\documentclass[aps,twocolumn,showpacs]{revtex4}
\usepackage{amsfonts}
\usepackage{amsmath}
\usepackage{amssymb}
\usepackage{graphicx}
\usepackage{fancyheadings}
\usepackage{times,xspace,color}
\usepackage{bm}
\usepackage{multirow}

\def\fS{{\sf{S}}}

\def\bbbc{{\mathchoice {\setbox0=\hbox{$\displaystyle\rm C$}\hbox{\hbox
to0pt{\kern0.4\wd0\vrule height0.9\ht0\hss}\box0}}
{\setbox0=\hbox{$\textstyle\rm C$}\hbox{\hbox
to0pt{\kern0.4\wd0\vrule height0.9\ht0\hss}\box0}}
{\setbox0=\hbox{$\scriptstyle\rm C$}\hbox{\hbox
to0pt{\kern0.4\wd0\vrule height0.9\ht0\hss}\box0}}
{\setbox0=\hbox{$\scriptscriptstyle\rm C$}\hbox{\hbox
to0pt{\kern0.4\wd0\vrule height0.9\ht0\hss}\box0}}}}

\newcommand{\ket}[1]{|{#1}\rangle}
\newcommand{\bra}[1]{\langle{#1}|}

\newcommand{\bk}{{\bf k}}

\preprint{}  
\begin{document}

\title{BCS-to-BEC crossover from the exact BCS solution}

\author{G. Ortiz$^{1}$ and J.~Dukelsky$^{2}$}
\address{$^{1}$ Theoretical Division, Los Alamos National Laboratory,
Los Alamos, New Mexico 87545, USA \\ $^{2}$ Instituto de Estructura de
la Materia, CSIC, Serrano 123, 28006 Madrid, Spain}

\date{Received \today }

\begin{abstract}

The BCS-to-BEC crossover, as well as the nature of Cooper pairs, in
superconducting and Fermi superfluid media is studied from the exact
ground state wavefunction of the reduced BCS Hamiltonian. As the
strength of the interaction increases, the ground state continuously
evolves from a mixed-system of quasifree fermions and pair resonances
(BCS), to pair resonances and quasibound molecules (pseudogap), and
finally to a system of quasibound molecules (BEC). A single unified
scenario arises where the Cooper-pair wavefunction has a unique
functional form. Several exact analytic expressions such as the binding
energy and condensate fraction are derived. We compare our results with
recent experiments in ultracold atomic Fermi gases.

\end{abstract}

\pacs{03.75.Ss, 02.30.Ik, 05.30.Fk, 74.20.Fg}
\author{}
\maketitle


The nature of Cooper pairs in the BCS-BEC crossover has regained
attention due to the observation of a large fraction of preformed
fermion pairs on the BCS side of the Feshbach resonance in ultracold
atomic Fermi gases \cite{Regal}. While several theoretical explanations
were proposed \cite{Ho}, the interpretations are still controversial.
The root of the controversy is understanding what represents a Cooper
pair in a correlated Fermi system, a concept not clearly defined in the
literature. This paper discusses this concept at the very foundational
level and proposes a qualitatively different scenario of the BCS-BEC
crossover, based on the exact solution to the BCS Hamiltonian
\cite{RichN}.
Only in the extreme BEC limit does this new scenario and Leggett's {\it
naive} ansatz \cite{leggett} become identical.

While the superconducting and Fermi superfluid thermodynamic states
represent a mixed-system of quasifree and pair-correlated fermions, the
molecular BEC which arises in the dilute and strong coupling limit has
all fermions  {\it bound} into pairs forming a unique macroscopic
quantum state. It is by now well accepted in which sense these states
represent a Bose-Einstein (BE) condensation. What defines a BE condensation
in an interacting $N$-particle system is spontaneous gauge
symmetry breaking, or phase coherence, of its quantum state
(a concept that strictly applies in the thermodynamic limit (TL)).
Yang \cite{Yang}
provided a detailed mathematical characterization of this phenomenon
which manifests itself as off-diagonal long-range order
(ODLRO) or, equivalently, by the existence of an eigenvalue of order $N$
in a reduced density matrix $\hat{\rho}_n$,
where $n$ is the number of particles forming the
{\it smallest unit}  that condenses.

The above definition of a
BEC does not imply the naive view of a many-body state of quantum
objects with identical internal wavefunctions. Indeed, we will
argue that a current-carrying {\it mean-field} ground state (GS) of a
uniform superconducting or Fermi superfluid $N$-particle system is of
the form
\begin{eqnarray}
\Psi(x_1,\cdots,x_N)=  {\cal A} \left [ \phi_1(x_1, x_2) \cdots
\phi_{N/2}(x_{N-1}, x_N) \right ] \ , \label{Var}
\end{eqnarray}
with $x_j=({\bf r}_j, \sigma_j)$, antisymmetrizer ${\cal A}$, and the {\it
pair} state
\begin{equation}
\phi_{\alpha}(x_i,x_j)=\
e^{i\mathbf{q}\cdot\frac{(\mathbf{r}_i+\mathbf{r}_j)}{2}} \
\varphi_\alpha(\mathbf{r}_i-\mathbf{r}_j)  \
\chi(\sigma_i,\sigma_j) \ ,
\end{equation}
where $\chi$ is a spin function of a certain symmetry, $\mathbf{q}$ is
the pair center-of-mass momentum, and $\varphi_\alpha(\mathbf{r})$ the
internal wavefunction which may represent either a quasimolecular
resonant state or a scattering state  (i.e., a mixed state), depending
upon the strength of the interaction between particles.

For pedagogical reasons we will recall the one-Cooper-pair problem and
then address the question of what happens when we add more pairs
($\mathbf{q}=\mathbf{0}$ in the GS). The Cooper-pair solution can be
recovered from the wavefunction (\ref{Var}) by assuming that $N-2$ fermions
$c^\dagger_{\bk \sigma}$ are free, filling the lowest momentum states
$\bk$ (of energy $\varepsilon_\bk$) up to the Fermi momentum $\bk_F$,
while {\it only} an additional fermion pair (with momenta $k> k_F$)
feels an attractive $s$-wave interaction in the spin singlet channel
\begin{equation}
\ket{\Psi}_{\sf C}=\sum_{k>k_{F}}\frac{1}{2\varepsilon_\bk-E}
c_{\bk\uparrow}^{\dagger}c_{-\bk\downarrow}^{\dagger} \ \ket{F} \ .
\label{Cpair}
\end{equation}
The role of the Fermi sea, $\ket{F}=\prod_{k\leq k_F}
c_{\bk\uparrow}^{\dagger}c_{-\bk\downarrow}^{\dagger}\ket{0}$, is
to Pauli-block states below the Fermi energy $\varepsilon_F$.
Assuming that the attractive pairing interaction $G<0$ is constant
around the Fermi energy, the eigenvalue $E$ is always negative
implying that the Cooper pair is bound regardless of the strength
of the attractive interaction. The Fermi sea is then unstable
against the formation of {\it bound pairs} of electrons.

What happens when the pairing interaction {\it also} affects electrons
in the Fermi sea? The answer to this question is the BCS theory whose
canonical form considers a state of the form (\ref{Var}) with all {\it
identical} internal wavefunctions
\begin{equation}
\hat{P}_M \ket{\sf BCS} =\left(  \Lambda^{\dagger} \right)^{M} \ket{0}
\ , \ \Lambda^{\dagger}=\sum_{\bk}\frac{v_{\bk}}{u_{\bk}}
c_{\bk\uparrow}^{\dagger} c_{-\bk\downarrow}^{\dagger} \ ,
\label{MFBCS}
\end{equation}
where $\hat{P}_M$ is the projector onto the state with $M$
pairs, and $v_\bk$, $u_\bk$ are the BCS amplitudes
$\binom{v_{\bk}^2}{u_{\bk}^2}=\frac{1}{2}\left
(1\mp\frac{\varepsilon_\bk-\mu}{
\sqrt{(\varepsilon_\bk-\mu)^2+\Delta^2}} \right )$ with $\Delta$
the gap and $\mu$ the chemical potential.
Since the structure of the BCS pair is averaged over the whole system,
it does not characterize a Cooper pair in the superconducting medium
except in the extreme strong-coupling and dilute limits
where all pairs are {\it bounded} and non-overlapping. Sometimes, the
structure of the Cooper pair is associated with the pair-correlation
function $\bra{\sf BCS}
c_{\bk\uparrow}^{\dagger}c_{-\bk\downarrow}^{\dagger }\ket{\sf BCS}
=u_{\bk}v_{\bk}$ describing the pair correlation among fermions of the
same pair as well as the exchange between fermions of different pairs.

What is the nature of a Cooper pair in a Fermi superfluid or
superconducting medium? To address this question we will  use
the exact solution to the reduced BCS Hamiltonian
\begin{eqnarray}
H &=&\!\!\sum_{\mathbf{k}}\varepsilon_{\mathbf{k}}\
{n}_{\mathbf{k}}+\frac{G}{V}\sum_{\mathbf{ k,k^{\prime }}}\
c_{\mathbf{k\uparrow }}^{\dagger }c_{-\mathbf{k\downarrow }
}^{\dagger }c_{-\mathbf{k^{\prime }\downarrow
}}^{\;}c_{\mathbf{k^{\prime }\uparrow }}^{\;}  \ ,
\label{BCS}
\end{eqnarray}
proposed by Richardson 40 years ago \cite{RichN,NB}. $H$
involves all terms with time-reversed
pairs ($\bk\uparrow,-\bk\downarrow$) from a contact
interaction. It is consistent with an effective single-channel
description of the BCS-BEC crossover theory \cite{leggett}
in terms of a zero-range potential. Realistic finite-range
interactions produce qualitatively similar results along the
crossover \cite{parish}.

For simplicity we will consider
$N=N_\uparrow+N_\downarrow$ spin-1/2 (i.e., 2-flavor) fermions
in a three-dimensional box of volume $V$ with periodic
boundary conditions, interacting through an attractive constant
($s$-wave-singlet-pairing) potential. (Other pairing symmetries
can also be accommodated.) Exact $N=2M+\nu$ particle eigenstates
of $H$ can be written as
\begin{equation}
\ket{\Psi}=\prod_{\ell=1}^M \fS^+_\ell \ket{\nu} \ , \mbox{with }
\fS^+_\ell=\sum_\bk \frac{1}{2 \varepsilon_\bk -E_\ell}
c_{\mathbf{k\uparrow}}^{\dagger}
c_{-\mathbf{k\downarrow}}^{\dagger} ,
\label{ansatz}
\end{equation}
where $\left\vert \nu\right\rangle \equiv$ $\left\vert
\nu_{1},\nu_{2}\cdots,\nu_{L}\right\rangle$ is a state of $\nu$ unpaired
fermions ($\nu=\sum_\bk\nu_\bk$, with $\nu_\bk=1,0$) defined by
$c_{-\mathbf{k\downarrow}}^{\;}c_{\mathbf{k\uparrow}}^{\;} \left\vert
\nu\right\rangle =0$, and $n_\bk \left\vert \nu\right\rangle
=\nu_{\bk}\left\vert \nu\right\rangle$. $L$ is the total number of
single particle states. The GS $\ket{\Psi_0}$ is in the $\nu=0$
($N_\uparrow\!=\!N_\downarrow$) sector.

Each eigenstate $\ket{\Psi}$ is completely defined by a set of $M$ (in
general, complex) spectral parameters (pair energies) $E_\ell$ which
are a solution of the Richardson's equations
\begin{equation}
1+\frac{G}{V}\sum_\bk\frac{1-\nu_\bk}{2\varepsilon_\bk-E_\ell}+
\frac{2G}{V}\sum_{m \left( \neq \ell \right) =1}^{M}\frac {1}{
E_{\ell}-E_{m}}=0 \ , \label{Rich}
\end{equation}
and the eigenvalues of $H$ are ${\cal E}=\sum_\bk \varepsilon_\bk
\nu_\bk+\sum_{\ell=1}^{M}E_{\ell}$ \cite{CO,NB}.
A crucial observation is that if a complex $E_\ell$ satisfies
(\ref{Rich}), its complex-conjugate $E^*_\ell$ is also a solution. Thus,
$\ket{\Psi}$ restores time-reversal invariance. The ansatz
(\ref{ansatz}) is a natural generalization of the Cooper-pair problem
without an inert Fermi sea, and with all pairs subjected to the pairing
interaction. The pair structure in (\ref{ansatz}) is  similar to the
Cooper pair in (\ref{Cpair}), and the many-body state has the form
(\ref{Var}) with all pairs {\it different}.

Since we are concerned with uniform bulk Fermi systems, we are
interested in the TL (i.e., $N, V
\rightarrow \infty$ with $\rho=N/V=k_F^3/3\pi^2$ = constant).
This limit was studied by Gaudin \cite{gaudin} when the energy
spectrum $\Omega \in [-\omega,\omega]$ is bounded, and $\nu=0$.
Eqs. (\ref{Rich}) reduce to the well-known {\it gap equation}
\begin{equation}
\frac{1}{2}\int_{\Omega}d\varepsilon ~\frac{g(\varepsilon)}{\sqrt{
\left( \varepsilon -\mu\right) ^{2}+\Delta^{2}}}+\frac{1}{G}=0 \ ,
\label{gaudin_gap}
\end{equation}
where $g(\varepsilon)$ represents the density of states. In our
case $\Omega \in  [0,\infty]$ is unbounded with
$g(\varepsilon)=m^{3/2}\varepsilon^{1/2}/(\sqrt{2}
\pi^{2}\hbar^{3})$ for $\varepsilon_\bk=\hbar^2k^2/2m$. Due to the
absence of an upper cutoff, Eq. (\ref{gaudin_gap}) is singular and
the TL becomes a subtle mathematical procedure whose solution will
be presented here \cite{bogo}.

\begin{figure}[htb]
\vspace*{-1.0cm}
\hspace*{-0.7cm}
\includegraphics[angle=0,width=10.0cm,scale=1.0]{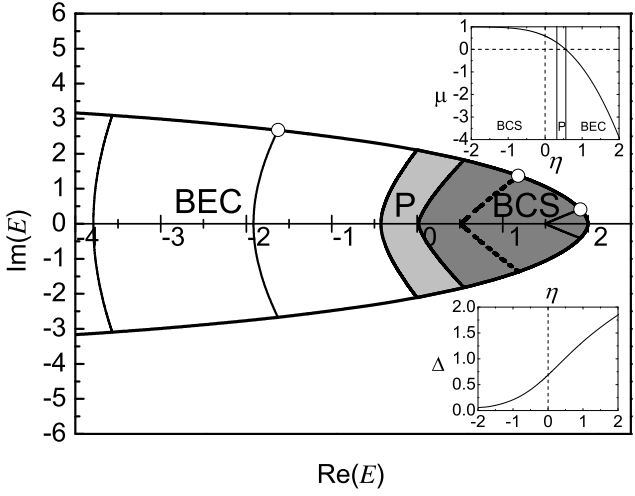}
\caption{Ground state crossover diagram displaying the different
regimes as a function of $\eta$. A few arcs $\Gamma$, whose extremes
are $E=2(\mu\pm i\Delta)$, are displayed. The dashed line corresponds
to the Feshbach resonance $\eta=0$. Open circles represent the values
of $E$ for which Cooper-pair wavefunctions are plotted in Fig.
\ref{fig3}. Insets show the behavior of the gap $\Delta$ and chemical
potential $\mu$. }
\label{fig1}
\end{figure}

\begin{table}[thb]
\begin{center}
\hspace*{-0.5cm}
\begin{tabular}{|c||c|c|c|c|}
\hline
$\eta$ & \hspace*{0.7cm}$\mu$\hspace*{0.7cm} & $\Delta$ & ${\cal E}_B$ &
$\lambda$\\
\hline $\rightarrow - \infty$ & $1+(\frac{\pi}{8}\eta
-\frac{5}{8}) \Delta^2$ & $\frac{8}{e^{2}} \exp{\frac{\pi
\eta}{2}}$ & $\frac{3}{4}\Delta^2$ & $\frac{3\pi}{e^{2}} \exp{\frac{\pi
\eta}{2}}$
\\
\hline $0$ & -$\frac{x}{E\left( \frac{1-x}{2}\right)^{\frac{2}{3}}}$ &
$\frac{\sqrt{1-x^2}}{E\left(
\frac{1-x}{2}\right)^{\frac{2}{3}}}$ & $\frac{6}{5} \left ( 1+ \frac{x}{E\left(
\frac{1-x}{2}\right)^{\frac{2}{3}}}\right )$ & $\frac{3\pi}{8\sqrt{2}}
\frac{\sqrt{1+x}(1-x)}{E\left(\frac{1-x}{2}\right)}$
\\
\hline $\frac{8\pi^{2/3}}{\Gamma[\frac{1}{4}]^{8/3}}$ & 0 &
$\sqrt{2\eta}$ & $\frac{6}{5} \left ( 1+ \frac{\pi
\eta^2}{4}\right )$ & $\frac{3\pi^{3/2}}{\sqrt{2}\Gamma[\frac{1}{4}]^{2}}$
\\
\hline $\rightarrow + \infty$ & $-\eta^2$ &
$\sqrt{\frac{16\eta}{3\pi}}$ & $2 \eta^2$ & 1
\\
\hline
\end{tabular}
\end{center}
\caption{Analytic expressions for selected values of $\eta=1/k_Fa_s$;
$x$ is the root of $P_{\frac{1}{2}}$, i.e. $P_{\frac{1}{2}}(x)=0$, and
$E(y)$ is the complete elliptic integral of the second kind.
Note that $5\pi^2{\cal E}_0(\eta=0)/k_F^3=\mu(\eta=0)\approx 0.590606$. }
\label{table1}
\end{table}
There are two ways to regularize the problem: One can either
introduce an energy cutoff in the integrals or one can cancel the
singularities with physical quantities whose bare counterpart
diverges in the same way \cite{leggett}. For this problem, Eq.
(\ref{BCS}), the physical quantity is the scattering length $a_s$
given by
\begin{equation}
\frac{m}{4\pi \hbar^2
a_s}=\frac{1}{G}+\frac{1}{2}\int_{0}^{\infty} d\varepsilon
~\frac{g(\varepsilon)}{\varepsilon} \ .
\end{equation}
The non-singular gap equation (after
integration \cite{papen}) is
\begin{eqnarray}
\frac{1}{k_F a_s}=\eta &=& \sqrt[4]{\mu^{2}+\Delta ^{2}} \
P_{\frac{1}{2}}\left( -\frac{\mu}{\sqrt{\mu^{2} +\Delta ^{2}}}\right)
\ ,
\label{gap}
\end{eqnarray}
where energies are now in units of
$\varepsilon_F=\hbar^2k_F^2/2m$ and lengths in units of
$\xi_F=1/k_F$. $P_\beta(x)$ is the Legendre function of the first
kind of degree $\beta$. The equations for the conservation of the
number of pairs $M$
\begin{eqnarray}
-\frac{4}{3\pi} \!\!&=&\!\!\eta \mu  +  (\mu^{2}+\Delta ^{2})^{3/4} \
P_{\frac{3}{2}}\left(  -\frac{\mu}{\sqrt{\mu^{2}+\Delta
^{2}}}\right) ,
\label{mu}
\end{eqnarray}
and GS energy density, for arbitrary strength $\eta$,
\begin{eqnarray}
{\cal E}_0&=&\frac{1}{V} \frac{\bra{\Psi_0} H
\ket{\Psi_0}}{\langle \Psi_0 | \Psi_0 \rangle}
=-\frac{k_F^3}{20\pi} \left [ \frac{\eta \Delta^2}{2} -
\frac{4}{\pi} \mu \right ] \ ,
\label{energy0}
\end{eqnarray}
do not need regularization, these are {\it exact} results. Indeed, for a
given $\eta$, the chemical potential $\mu$ and gap $\Delta$ need to be
determined self-consistently from Eqs. (\ref{gap}) and (\ref{mu}) (Fig.
\ref{fig1}). Then, the GS energy can be computed as a function of
density using Eq. (\ref{energy0}). It shows no phase segregation. The
{\it exact} binding (or condensation) energy per electron pair ${\cal
E}_B$ is  (see Table \ref{table1})
\begin{eqnarray}
{\cal E}_B&=&\frac{3 \pi}{10} \left [\frac{\eta \Delta^2}{2}- \frac{4}{\pi}
(\mu-1) \right ] \ .
\label{energyb}
\end{eqnarray}

The complete solution of Eqs. (\ref{Rich}) in the TL \cite{gaudin}
amounts to determining (for a given $\eta$) the set of pair
energies $E_\ell$ which, for the GS,  form a single open,
continuous, and reflection-symmetric arc $\Gamma$ with extreme
points $E_F=2(\mu \pm i \Delta)$. Here, we simply present the
results without derivation. The equation for $\Gamma$ is

\begin{eqnarray}
0&=&\operatorname{Re} \left[ \int_{0}^{\infty }d\varepsilon
~\sqrt{\varepsilon }\left( \frac{z+\left( \varepsilon -\mu
\right)\ln \left[ \frac{\left( E -\mu \right) +z}{i \Delta
}\right]}{\sqrt{\left( \varepsilon -\mu \right) ^{2}+\Delta^{2}}}
- \right .\right . \nonumber
\\
&&\hspace*{-1.2cm} \left. \left. \ln \left[ \frac{ \Delta
^{2}+\left( E -\mu \right) \left( \varepsilon -\mu \right) +z \
\sqrt{\left( \varepsilon -\mu \right) ^{2}+\Delta ^{2}}}{ i\Delta
\left( \varepsilon -E \right)} \right] \right ) \right]
\end{eqnarray}
where $z=\sqrt{\left( E -\mu \right) ^{2}+\Delta ^{2}}$.

Figure \ref{fig1} shows the BCS-BEC crossover diagram and several
arcs corresponding to particular values of $\eta$. Since a
crossover diagram is not a phase diagram, i.e. there is no
symmetry order parameter or non-analyticities sharply
differentiating the regions,  boundaries are in principle
arbitrary.

Here we have adopted the following criteria: The geometry of the
arcs $\Gamma$ serves us to establish a criterium for defining
boundary regions in the crossover diagram.
In the extreme weak-coupling limit, $\eta \rightarrow -\infty$,
the pair energies are twice the single particle energies ($E_\ell
\rightarrow 2 \varepsilon_\bk$). As soon as the interaction is
switched on, i.e. $k_Fa_s$ is an infinitesimal negative number, a
fraction $ {\sf f}$ of the pair energies close to $2\mu$ become
complex, forming an arc $\Gamma$ in the complex plane. The
fraction $1-{\sf f}$ of fermion pairs below the crossing
energy $2 \varepsilon_c$ of $\Gamma$ with the real axis have real
pair energies. They are still in a sea of uncorrelated pairs with
an effective Fermi energy $\varepsilon_c$.  The dark grey region
labelled BCS, which extends from  $\eta=-\infty$ to $\eta=0.37$,
is characterized by a mixture of  complex pair energies with a
positive real part and real pair energies. $\eta=0.37$ is
the value at which all pair energies are complex, i.e.
${\sf f}=1$, and the effective Fermi sea has
disappeared. Within the BCS region we plotted the
arcs for $\eta=-1$ (solid line), with a fraction  ${\sf f}=0.35$, and
$\eta=0$ (dashed line) with ${\sf f}=0.87$.  The {\it pseudogap} region
P, indicated in light grey, extends from $\eta=0.37$ to $\eta=0.55$
where $\mu=0$. Within this region the real part
of the pair energies changes from positive to negative, and P describes a
mixture of Cooper resonances and quasibound molecules. The BEC (white)
region, $\eta>0.55$, is characterized by all pair energies having
negative real parts, i.e. all pairs are quasibound molecules. As $\eta$
increases further, $\Gamma$ tends to an almost vertical line with
$\operatorname{Re}(E)\sim 2\mu$, and
$ -2\Delta \leq \operatorname{Im}(E) \leq 2\Delta$.

If the literature is not clear about
the size $\xi$ of a Cooper pair, it is partly because it is not
clear what a Cooper pair is. Pippard, in his nonlocal electrodynamics
of the superconducting state, introduced the characteristic length
$\xi_0$ by using an uncertainty-principle argument involving only
electrons within a shell of width 2$\Delta$ around the Fermi energy.
The resulting quantity, named Pippard's coherence length, $\xi_0=2/(\pi
\Delta)$, is usually associated to $\xi$. On the other hand, an
acceptable definition could be
$\xi=\sqrt{\bra{\varphi} r^2 \ket{\varphi}} \ , \ \langle
\varphi | \varphi \rangle =1$.
From Eq. (\ref{ansatz}), the Cooper-pair wavefunction is
\begin{equation}
\varphi_E(\mathbf{r})=\frac{1}{V}\sum_{\mathbf{k}} \varphi_\bk^E \
e^{i\mathbf{k}\text{\textperiodcentered}\mathbf{r}}=
A \ \frac{e^{-r\sqrt{-E/2}}}{r} \ , \label{Cooper}
\end{equation}
with $A^2=\operatorname{Im}(\sqrt{E/2})/2\pi \xi_F^3$,
$\varphi_\bk^E=C/(2\varepsilon_{\bf k}-E)$, and $C$ being a normalization
constant. Thus,
$\xi_E=1/\operatorname{Im }(\sqrt{E})$.
In the weak-coupling BCS limit ($\Delta \ll \mu \approx 1$),
 when $E=E_F$, we get
$\xi_E=\pi\xi_0/\sqrt{2}$. On the other hand, in the same limit, if one
uses $\varphi_\bk^{\sf P}= C_{\sf P} u_\bk v_\bk$, one gets $\xi_{\sf
P}=\xi_E/2$, and if one uses $\varphi_\bk^{\sf BCS}= C_{\sf BCS}
v_\bk/u_\bk$, one gets $\xi_{\sf BCS}=\sqrt{21/2}$.

\begin{figure}[htb]
\hspace*{-0.7cm}
\includegraphics[width=8.0cm,scale=1.]{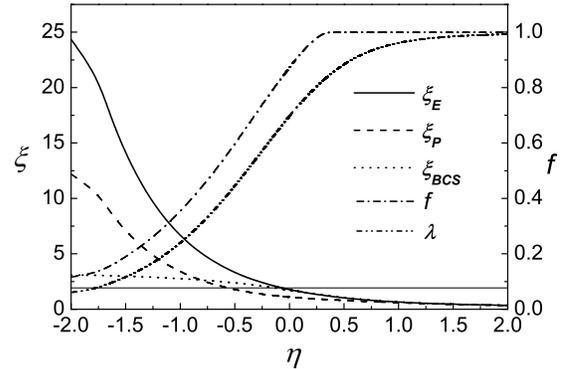}
\caption{Different definitions of Cooper-pair sizes $\xi$, and
condensate fraction $\lambda$ and ${\sf f}$ as a function of the
interaction strength $\eta$. The horizontal solid line represents the
interparticle distance $r_s=\sqrt[3]{9\pi/4}$. }
\label{fig2}
\end{figure}
Analytic expressions (using dimensional regularization) for
$\xi_{\sf P}$ and $\xi_{\sf BCS}$ can also be obtained for
arbitrary coupling strength. Figure \ref{fig2} displays the
different sizes, all labelled by $\xi$, as a function of $\eta$.
Clearly, in (\ref{ansatz}) there is no unique pair size $\xi_E$ but
a distribution of sizes; Fig. \ref{fig2} shows in solid line the
smallest size in $\Gamma$ corresponding to $E_F$.
On the other hand, there is a unique pair size for $\varphi_{\sf
P}$ and $\varphi_{\sf BCS}$; the dashed and dotted lines correspond
to the $\xi_{\sf P}$ and $\xi_{\sf BCS}$ sizes respectively.
Notice the relation between sizes and the interparticle distance
$r_s=\sqrt[3]{9\pi/4}$ which is indicated as a full horizontal
line. While $\xi_E$ and $\xi_{\sf P}$ increase for negative $\eta$
values, and eventually diverge when $\eta \rightarrow -\infty$,
$\xi_{\sf BCS}$ tends to a constant value, related to the
interparticle distance, showing its essentially uncorrelated
nature. The fact that $\xi_{\sf P} < \xi_E/2$ in the overlapping
BCS region is a clear demonstration that $\xi_{\sf P}$ measures
the mean distance between time-reversed pairs irrespective of the
Cooper pair they belong to. For non-overlapping pairs (BEC region)
both sizes converge to the same values.


\begin{figure}[htb]
\hspace*{-0.7cm}
\includegraphics[width=8.0cm,scale=1.]{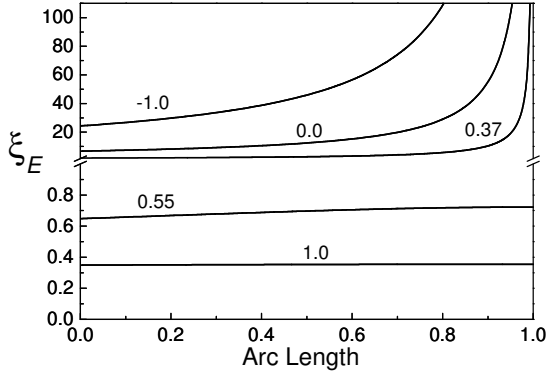}
\caption{Cooper-pair sizes $\xi_E$ along the arcs $\Gamma$ for the
values of $\eta$ depicted in Fig. 1.} \label{fig3}
\end{figure}

Following Yang \cite{Yang}, ODLRO in
$\hat{\rho}_2({\bf r}_1\!\!\uparrow\!{\bf r}_2\!\!\downarrow \!\!| \, {\bf
r}_3\!\!\uparrow\!{\bf r}_4\!\!\downarrow) \rightarrow \phi^*({\bf
r}_1\!\!\uparrow,{\bf r}_2\!\!\downarrow)  \phi({\bf
r}_3\!\!\uparrow,{\bf r}_4\!\!\downarrow)$
may be used to define

\begin{equation} \lambda =
\frac{2}{N} \!\!\int \!\!d {\bf r}_1 d {\bf r}_2\ | \phi({\bf
r}_1\!\!\uparrow,{\bf r}_2\!\!\downarrow)|^2=\frac{3 \pi}{16}
\frac{\Delta^2}{\operatorname{Im}(\sqrt{\mu+ i \Delta})}
\end{equation}
as a measurement of the condensate fraction. Figure \ref{fig2} shows
$\lambda$ and the fraction ${\sf f}$ of (Cooper) pairs in the arc,
that is, the fraction of complex spectral parameters.
Although $\lambda$ can qualitatively describe the fraction of
correlated pairs, the fraction ${\sf f}$ gives
the value $1$ at the BCS-pseudogap boundary ($\eta=0.37$) while
$\lambda=1$ for $\eta \rightarrow \infty$. We
note that at resonance ($\eta=0$), ${\sf f}=87\%$ of the fermions
form Cooper pairs ($\lambda \approx 0.7$). These results provide a
qualitative explanation of the experiments in \cite{Regal} without
resorting to a projection method \cite{Ho}. Close to resonance on
the BCS side, a fraction ${\sf f}\sim 80\%$ of the atoms form
Copper pairs with sizes comparable to $r_s$. Those atom pairs are
efficiently transformed into {\it quasimolecules} during a rapid
magnetic field ramping across the resonance giving rise to the
molecular condensate fractions observed experimentally.

One may argue that selecting the smallest pair from each arc to compare
(in Fig.\ref{fig2}) Cooper-pair sizes along the crossover might not be
representative of the Cooper-pair distribution within each arc. In Fig.
\ref{fig3} we show the internal variations of the pair sizes within the
arcs. Although in the BCS region there is a distribution of sizes from
the smallest pair with $E=E_F$ in the extremes of the arc to an almost
infinite size pairs when they are close to the real axis, already at
resonance ($\eta=0$) most of the pairs have the same structure. This
latter feature becomes more pronounced in the BEC region.

\begin{figure}[htb]
\hspace*{-0.7cm}
\includegraphics[angle=0,width=7.0cm,scale=1.0]{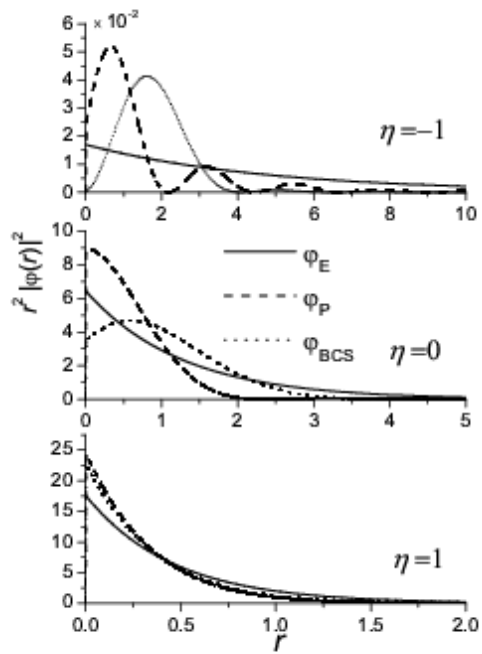}
\caption{Cooper-pair wavefunctions for different $\eta$. The upper
and middle (Feshbach resonance) panels correspond to the BCS
region, while the bottom one is in the BEC region. Except for the
Cooper $\varphi_E$ case, the other two wavefunctions always vanish
at $r=0$. It is only in the limit $\eta \rightarrow + \infty$ that
the three states {\it exactly} coincide.} \label{fig4}
\end{figure}

The various definitions of Cooper-pair wavefunctions are depicted in
Fig. \ref{fig4}, which compares $\varphi_{\sf BCS}(r)$, $\varphi_{\sf
P}(r)$, and $\varphi_{E}(r)$ for interaction strengths which correspond
to the BCS, Feshbach resonance, and BEC regions of Fig. \ref{fig1}.
Notice that while $\varphi_{\sf BCS}$, and $\varphi_{\sf P}$ are highly
oscillatory in the weak-coupling region, this is not the case with
$\varphi_{E}$ which always has an exponential form. Clearly, a single
and unified picture emerges for the crossover when using a many-body
state such as (\ref{Var}): $\varphi_{E}$ evolves
smoothly through the crossover as it should. It is important to
mention that the three wavefunctions are {\it exactly} the same in
the strong coupling limit $\eta \rightarrow +\infty$.
In this limit the {\it naive} ansatz of Leggett \cite{leggett} and
the GS (\ref{Var}) coincide, becoming a simple Pfaffian state.


In summary, we studied the BCS-BEC crossover problem, as well as the
nature of Cooper pairs in a correlated Fermi system, from the exact GS
$\ket{\Psi_0}$ of the reduced BCS Hamiltonian. We have analytically
determined its exact TL for the quadratic single-particle dispersion,
and calculated several quantities of physical interest, such as the
binding (or condensation) energy and the condensate fraction. The
validity of the present description is not restricted to integrable
pairing Hamiltonians, but rather the ansatz $\ket{\Psi_0}$, which  is a
natural extension of the Cooper problem, could be used as a variational
state for more realistic interactions. The Cooper-pair wavefunction
$\varphi_E$ has a fixed functional $s$-wave form that accommodates
pair-correlated fermions close to the Fermi energy in the weak coupling
limit as well as quasibound molecules in the BEC limit. It is free from
the characteristic oscillations displayed by $\varphi_{\sf BCS}$ and
$\varphi_{\sf P}$ related to exchange among pairs. The BCS region in the
crossover diagram represents a mixture of Cooper pairs and quasifree
fermions.
It may be argued that the single-channel model is insufficient to
describe the system close to resonance where the degrees of freedom
associated to the molecular closed channel should be explicitly
incorporated. A derivation analogous to the one presented here
can be pursued by using a recently proposed atom-molecule integrable model
\cite{duke} which captures the essential features of the two-channel
model \cite{HoTi}. The structure of the Cooper
pair in this new model is the same as in (\ref{ansatz}).

We acknowledge discussions with G. G. Dussel, J. Gubernatis, A. J. Leggett, D.
Loss, G. Sierra, E. M. Timmermans, and S. A. Trugman. JD
acknowledges financial support from the Spanish DGI under grant
 BFM2003-05316-C02-02.

\end{document}